\documentclass[twocolumn,tighten,times,tracking, twocolappendix]{aastex631}

\usepackage{amsmath}
\usepackage{color}
\usepackage{longtable}
\usepackage{ulem,array}
\usepackage{grffile}
\usepackage{comment}
\usepackage[dvipsnames]{xcolor}

\usepackage{array}
\usepackage{tabularx} 
\newcolumntype{Y}{>{\centering\arraybackslash}X} 
\newcolumntype{M}[1]{>{\centering\arraybackslash}m{#1}}
\newcolumntype{N}{@{}m{0pt}@{}}

\usepackage{xpatch}
\usepackage{hyperref}
\hypersetup{
    pageanchor=false,
	colorlinks=true,
	breaklinks=true,
	citecolor=blue,
	allcolors=blue,
	frenchlinks=true
}

\makeatletter
\xpatchcmd\NAT@citex
{%
	\@citea\NAT@hyper@{%
		\NAT@nmfmt{\NAT@nm}%
		\hyper@natlinkbreak{\NAT@aysep\NAT@spacechar}{\@citeb\@extra@b@citeb}%
		\NAT@date
	}%
}
{%
	\@citea
	\NAT@nmfmt{\NAT@nm}%
	\NAT@aysep\NAT@spacechar
	\NAT@hyper@{\NAT@date}%
}
{}{}
\xpatchcmd\NAT@citex
{%
	\@citea\NAT@hyper@{%
		\NAT@nmfmt{\NAT@nm}%
		\hyper@natlinkbreak{\NAT@spacechar\NAT@@open\if*#1*\else#1\NAT@spacechar\fi}%
		{\@citeb\@extra@b@citeb}%
		\NAT@date
	}%
}
{
	\@citea
	\NAT@nmfmt{\NAT@nm}%
	\NAT@spacechar\NAT@@open\if*#1*\else#1\NAT@spacechar\fi
	\NAT@hyper@{\NAT@date}%
}
{}{}
\makeatother

\newcommand{\mathbfit}[1]{\textbf{\textit{#1}}}
\renewcommand{\vec}[1]{\mathbfit{#1}}
\newcommand{\bs}[1]{\boldsymbol{#1}}
\newcommand{\alf}{Alfv\'en}

\newcommand{\vA}{v_{\rm A}}

\usepackage{multirow}
\usepackage{amssymb}
\usepackage[mathscr]{euscript}

\DeclareSymbolFont{matha}{OML}{txmi}{m}{it}
\DeclareMathSymbol{\varv}{\mathord}{matha}{29}

\SetSymbolFont{symbols}{bold}{OMS}{cmsy}{b}{n}
\DeclareSymbolFont{bmisymbols}{OML}{cmm}{b}{it} 
\DeclareMathSymbol{\bvarv}{0}{bmisymbols}{"1D}

\begin{document}

\author[0000-0001-9625-5929]{Mohamad Shalaby}
\affiliation{Perimeter Institute for Theoretical Physics, 31 Caroline St.~N., Waterloo, Ontario N2L 2Y5, Canada}
\affiliation{Waterloo Centre for Astrophysics, University of Waterloo, Waterloo, ON N2L 3G1, Canada}
\affiliation{Department of Physics and Astronomy, University of Waterloo, Waterloo, ON, N2L 3G1, Canada}
\affiliation{Horizon AstroPhysics Initiative (HAPI) Fellow}

\author[0000-0002-4683-8517]{Rouven Lemmerz}
\affiliation{Department of Physics, University of Wisconsin-Madison, Wisconsin 53706, USA}

\author[0000-0002-7275-3998]{Christoph Pfrommer}
\affiliation{Leibniz-Institut f{\"u}r Astrophysik Potsdam (AIP), An der Sternwarte 16, 14482 Potsdam, Germany}

\title{Saturation Mechanism of Cosmic Ray Streaming Instabilities with a Shell Distribution
}

\shorttitle{Saturation of CR Streaming Instability via Gyro-phase Bunching}

\shortauthors{Shalaby et al.}

\begin{abstract}
Cosmic rays (CRs) drive galactic winds and regulate galaxy growth while contributing to heating the central cooling plasma in dense galaxy clusters, making the nonlinear saturation of the instabilities that govern their transport a problem of central importance in astrophysics. Using fully kinetic particle-in-cell (PIC) simulations, we find that the streaming instability driven by CRs with a shell momentum distribution saturates through gyro-phase bunching around driven {\alf} waves. Fluid-PIC (FPIC) simulations with both ideal and Landau closures reproduce the PIC saturation amplitude,
and the same gyro-phase bunching mechanism, indicating that nonlinear Landau damping (NLLD) does not determine the saturation level at our simulation parameters ($\vA = 0.01c$, $n_{\rm CR}/n_i = 0.01$). Notably, the CR ions fully isotropize in the {\alf}-wave frame in all simulations, including the FPIC run with an ideal closure in which NLLD is entirely absent, showing that this isotropization does not require NLLD. We additionally confirm, using the FPIC ideal closure, that this saturation amplitude is unchanged across a tenfold increase in domain size, indicating that the mechanism is local rather than dependent on domain-scale processes. Together, these results indicate that, at least in this regime, NLLD is not the dominant saturation mechanism for the CR streaming instability, in tension with common assumptions built into CR transport models used in galaxy formation and interstellar medium simulations. Whether this conclusion extends to the lower CR densities and {\alf} speeds characteristic of the interstellar medium remains an open question that we address in an upcoming work.
\end{abstract}

\section{Introduction}
\label{sec:intro}

Cosmic rays (CRs) constitute a major energy component of the interstellar medium (ISM), with energy densities comparable to or exceeding those of thermal gas, magnetic fields, and turbulence \citep{Boulares1990,Ferriere2001,Zweibel2017}. They influence the heating, chemistry, and dynamics of the ISM over timescales of \(10^3\)--\(10^8\) years, and play a central role in galaxy formation and evolution \citep{Ruszkowski2023} by driving galactic winds from the multiphase interstellar medium \citep{2016Girichidis,2018Girichidis,2016Simpson,Sike2025,Kim2026}, forming galactic super-winds that regulate star formation on global scales \citep{2012Uhlig,Booth2013,Salem2014,2016PakmorIII,Ruszkowski2017,Thomas2025,Girichidis2024}, thereby providing pressure support to the circumgalactic medium \citep{2020Buck,Ji2020,Hopkins2021,Bieri2026}, which may modify the thermal instability \citep{Butsky2020,Weber2026}. Accelerated primarily in supernova remnants \citep{Bell1978a,Blandford1978}, CRs propagate through the Galaxy, scattering off magnetic fluctuations and eventually escaping after a few million years \citep{Ginzburg1964}. Their transport is thus a key ingredient in astrophysical systems ranging from the heliosphere to galaxy clusters \citep{Fisk1998,Zweibel2013,jacob2017a,jacob2017b,2018Jacob,Amato2018,Engelbrecht2022,Rankin2022}.

CRs stream along magnetic field lines, and when their drift velocity exceeds the {\alf} speed, they excite a family of instabilities that amplify magnetic fluctuations and scatter the CRs themselves \citep{Kulsrud1969}. The most widely studied of these is the gyro-resonant CR streaming instability (CRSI), which excites {\alf} waves at the ion gyroscale and forms the basis of the canonical self-confinement picture \citep{Wentzel1974,Skilling1975a,Skilling1975b,Skilling1975c}. More recently, a class of instabilities operating on intermediate scales between the ion and electron gyro-radii has been discovered \citep{sharp2,Shalaby2023}; driven by CRs with finite pitch angles, these excite whistler and electron-cyclotron waves that grow much faster than the CRSI \citep{sharp2,Shalaby2023,Lemmerz2023}.
The CRSI remains the primary mechanism for CR self-confinement in the ISM, but the intermediate-scale instability may also play an important role in certain regimes \citep{sharp2,Shalaby+2022ApJ}; both share a common resonant origin and can be understood within a unified framework \citep{Shalaby2023,Lemmerz2025}. Through their amplification of magnetic fields and subsequent scattering of CRs, these instabilities form a self-regulating feedback loop that sets the effective transport speed of CRs \citep{Bai2019,Zweibel2017,sharp2}.

The efficiency of CR transport depends critically on the saturation amplitude of these instabilities, i.e., the level of magnetic field fluctuations $\delta B/B_0$ at which they saturate \citep{Zweibel2017}. While the linear growth of the gyro-resonant instability is well understood \citep{Kulsrud1969,Wentzel1974}, the nonlinear saturation mechanism remains debated. A widely used assumption is that the instability saturates when growth is balanced by ion-neutral damping \citep{Kulsrud1969} or in the absence of neutrals by NLLD \citep{Lee1973,Miller1991}, in which two {\alf} waves of slightly different wavenumbers combine to form a beat wave that is damped by Landau damping and thus heats the background plasma. In the two-moment CR-magnetohydrodynamic (CRMHD) framework, NLLD is included as a key dissipation channel that regulates the CR--wave interaction and limits both wave growth and CR scattering \citep{Jiang2018,timon2019}, an assumption commonly adopted in subgrid CR transport models for galaxy formation simulations \citep{Thomas2026,Hopkins2026}.

Recent numerical work has advanced our understanding of the nonlinear evolution of the streaming instability \citep{2018Lebiga,Bai2019,Holcomb2019,sharp2,Lemmerz2025,Schroer2025}.
The MHD-PIC method is used by \citet{Bai2019} to study growth and saturation in a regime close to realistic ISM conditions ($n_{\rm CR}/n_i\sim 10^{-9}$, $\vA/c\sim 10^{-5}$), capturing the linear growth and quasi-linear diffusion of CRs and attributing saturation to quasi-linear diffusion and phase randomization across periodic boundaries; however, their fluid treatment of the background does not capture its full kinetic response, including NLLD.
\citet{Schroer2025} performed hybrid-PIC simulations and claimed that NLLD plays an important role in saturation, reducing the CR drift speed through scattering while keeping it super-{\alf}ic due to the damping of the generated waves. Conventional full-PIC simulations \citep{Holcomb2019,sharp2} have been limited by the extreme scale separation between thermal-plasma and CRs, as well as by low CR anisotropy, making the astrophysically relevant regime difficult to reach. Fluid-PIC (FPIC) simulations \citep{Lemmerz2023,Lemmerz2025} offer a middle ground, treating the background as a fluid while retaining a kinetic treatment of the CRs, enabling controlled comparison with fully kinetic simulations.
To date, no fully kinetic simulations have been systematically compared the saturation of the streaming instability with and without NLLD or directly identified the microscopic saturation mechanism for CR ions (CRi) with a shell distribution -- a gap this work addresses with a direct, controlled test at the parameters of our simulations.

In this work, we perform a fully kinetic PIC simulation of the CR streaming instability driven by CRs with a shell momentum distribution -- a limiting representation of a power-law CR population with a spectral index $p \to \infty$, for which the number density is dominated by particles near the minimum momentum $p_{\rm min}$. Since the growth of the instability is driven primarily by these low-energy particles, the shell distribution captures essential physics while avoiding the cost of resolving a broad momentum range \citep{Schroer2025}. We complement this with FPIC simulations using both ideal and Landau fluid closures for the background species \citep{Lemmerz2023}, which together isolate the role of NLLD: the PIC simulation includes NLLD self-consistently, with no assumption about its strength; the FPIC simulation with ideal closure omits Landau damping and consequently NLLD entirely by construction; and the FPIC simulation with Landau closure approximates linear Landau damping and NLLD via a fluid closure that, while validated against kinetic NLLD in related setups \citep{Lemmerz2023}, might underestimate the damping rate relative to a fully kinetic treatment. The PIC-vs-ideal-closure comparison is therefore our primary diagnostic of the role of NLLD, since it does not depend on the fidelity of any damping approximation; the Landau-closure simulation serves as a corroborating check.

We perform all simulations at $\vA = 0.01\,c$, $n_{\rm CR}/n_i = 0.01$, and background plasma-$\beta = 0.02$. Our main results are threefold. First, the PIC and FPIC ideal simulations agree closely on the saturation amplitude, $\delta B/B_0 \approx 0.25$, with the FPIC-Landau simulation reproducing the same value; using the FPIC ideal closure, we further show this amplitude is unchanged across a tenfold increase in domain size, indicating a local rather than domain-scale-dependent mechanism. Second, we identify the saturation mechanism as gyro-phase bunching around the driven {\alf} waves -- a local process that naturally saturates the instability, in contrast to simulations that artificially remove bunching via phase randomization and thereby likely overestimate the saturation level \citep[see, e.g.,][]{Bai2019}. Third, the close agreement between PIC and FPIC-ideal, in which no background damping channel is present, indicates that NLLD does not set the saturation level at our parameters, in tension with recent claims by \citet{Schroer2025}.
This comparison is extended by \citet{Lemmerz2026} to FPIC simulations with ideal and Landau closures at parameters closely matching those of \citet{Schroer2025}, and finds that the two closures continue to agree; because this comparison lacks a fully kinetic PIC benchmark, it corroborates rather than independently establishes the irrelevance of NLLD in that regime.
Together, these results indicate that, at least in the regime directly tested with fully kinetic PIC simulations, NLLD is not the dominant saturation mechanism assumed in the two-moment CRMHD framework, and that this conclusion plausibly extends to the realistic parameters.

This paper is organized as follows. Section~\ref{sec:setup} describes our numerical methods, simulation parameters, and initial CR distribution. Section~\ref{sec:results} presents our results: the growth of magnetic-field energy, the helicity decomposition of the driven waves, the identification of gyro-phase bunching as the main saturation mechanism, and the insignificance of NLLD. Section~\ref{sec:discussion} discusses implications for CR transport modeling, comparing with \citet{Bai2019,Schroer2025,timon2019}. We conclude in Section~\ref{sec:conclusions}.

\section{Numerical Simulations}
\label{sec:setup}

\begin{figure}
    \includegraphics[width=1.0\linewidth]{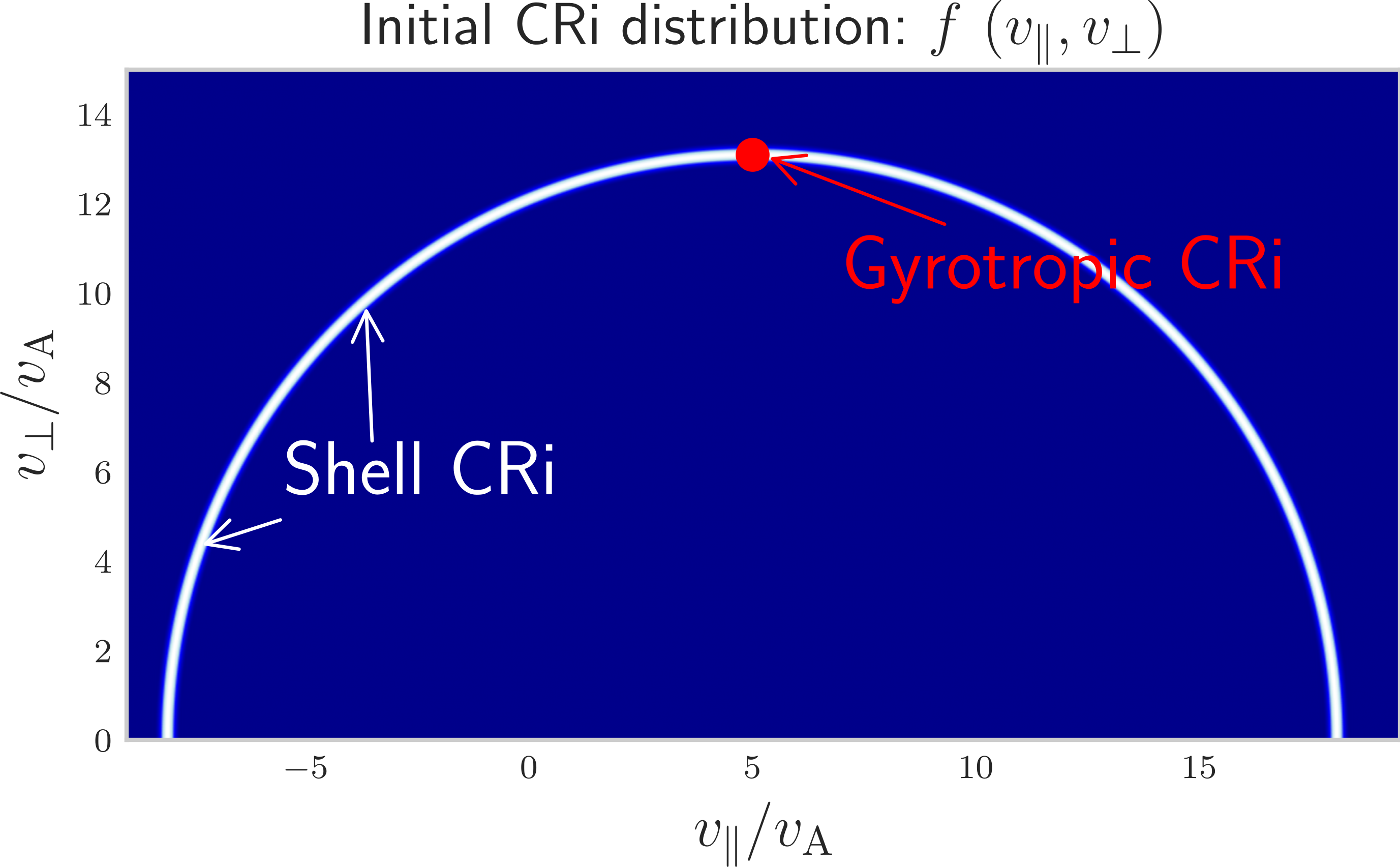}
    \caption{Visualization of the initial CRi velocity distributions. The gyrotropic (red) distribution was used in \citet{sharp2,Lemmerz2025}, while the shell (white) distribution is used in this work.}
    \label{fig:CRi}
\end{figure}

We perform 1D3V (one spatial and three momentum dimensions) simulations using the code SHARP-1D3V \citep{sharp,sharp2,Lemmerz2023}. 
The SHARP code avoids excessive numerical heating, conserves momentum exactly, and has been extensively validated for beam-plasma instabilities \citep{resolution-paper,sim_inho_18,th_inho_20}, CR-driven instabilities \citep{sharp2,Lemmerz2023,Lemmerz2025}, and electron-ion shock formation \citep{Shalaby+2022ApJ,Shalaby2024ApJL,shalaby2025a}.
We perform two classes of simulations: fully kinetic particle-in-cell (PIC) simulations, in which all species (background and CRs) are treated kinetically, and FPIC simulations, in which the background electrons and ions are treated as charged fluids while the CRs remain kinetic \citep{Lemmerz2023,Lemmerz2025}. In the FPIC simulations, we employ either an ideal gas closure or a Landau fluid closure for the background species \citep{Lemmerz2023}. The Landau closure employed is the $R_{31}$ closure by \citet{hunana2018}, which is evaluated in Fourier space at every time step and thereby recovers the kinetic response accurately over a large range of $k$.

All simulations are initialized with four spatially uniform species: background electrons and ions (both with density $n_i$), and CR electrons and ions (CRe and CRi, both with density $n_{\rm CR}$). We adopt a realistic ion-to-electron mass ratio $m_i/m_e = 1836$ and impose a uniform magnetic field $\bs{B}_0 = B_0 \bs{\hat{x}}$, giving an {\alf} speed $\vA = B_0/\sqrt{\mu_0 m_i n_i} = 0.01\,c$. To reduce computational cost while retaining the essential physics, we use an enhanced CR density contrast $\alpha \equiv n_{\rm CR}/n_i = 10^{-2}$ and an enhanced {\alf} speed $\vA = 0.01\,c$ in comparison to realistic interstellar medium parameters; these choices increase the instability growth rate and reduce the scale separation between the ion gyroradius and the electron skin depth, allowing us to capture the relevant growth timescales with fewer time steps. In the PIC simulation, the background species are initialized with Maxwellian distributions at $k_{\rm B} T_e = k_{\rm B} T_i = 10^{-6} m_i c^2$ and are at rest in the simulation frame; in the FPIC simulations, the electron and ion background fluids are initialized with the same temperatures.
Therefore, the background plasma-$\beta = 2 (v_\mathrm{th,i}/\vA)^2 = 0.02$. Even though NLLD by the background ions is expected to be small, because the ion thermal velocity is a tenth of the Alfv\'en speed, the electron thermal velocity is about $v_\mathrm{th,e}\approx 4.2 \vA$ and is therefore in principle able to provide efficient damping in kinetic theory.

The CR ions are initialized with a shell distribution in momentum space (see the white arc in Figure~\ref{fig:CRi}). In the CRi rest frame, their pitch-angle cosine $\mu \equiv v_\parallel / \sqrt{v_\parallel^2 + v_\perp^2}$ is uniformly distributed over $\mu \in [-1,1]$, and their total momentum magnitude is fixed to $v_0 = 13.1\,\vA$. This implies a uniform distribution of parallel velocities in the CR rest frame. We then apply a boost along $\bs{B}_0$ to impose a net drift velocity $v_{\rm dr} = 5\,\vA$. CR electrons are initialized with no momentum spread and are all drifting along $\bs{B}_0$ with speed $v_{\rm dr} = 5\,\vA$, which neutralizes the CRi current. This shell distribution is an idealization of the common model of the power-law CRi population with a spectral index $p \geq 4$.
For such a case, the number density is dominated by particles near the minimum momentum $p_{\rm min}$. Since the growth of the streaming instability is driven primarily by these low-energy particles, the shell distribution captures the essential physics while avoiding the computational cost of resolving a broad momentum range.

For the PIC simulations, background ions and electrons are represented by 7500 macroparticles per cell and per species; CR ions and electrons use 75 macroparticles per cell and per species \citep{sharp,sharp2}. For the FPIC simulations, the background species are treated as fluids, while the CR ions and electrons remain kinetic with the same macroparticle numbers \citep{Lemmerz2023}.
The simulation box is periodic, and all simulations are performed in the background plasma rest frame.

To investigate the saturation mechanisms and the potential role of NLLD, we perform simulations with domains that extend over multiple fastest-growing wavelengths. The fastest-growing wavelength, $\lambda_m$, is given by \citep{Zweibel2017} $\lambda_m = 2\pi ( (v_0 + v_{\rm dr})/\vA - 1 ) d_i \simeq 108\,d_i$, where $d_i \equiv c/\omega_{pi}$ is the ion skin depth and $\omega_{pi} = \sqrt{n_i e^2/\epsilon_0 m_i}$ is the ion plasma frequency.
We resolve the ion skin depth using $10 \times \sqrt{m_i/m_e} \sim 428$ computational cells.
We run a series of $5\lambda_m$ PIC and FPIC simulations with a domain size of $L = 583\,d_i \simeq 5.42\,\lambda_m$, and a $50\lambda_m$ FPIC simulation with ideal fluid closure, where $L = 5830\,d_i \simeq 54.2\,\lambda_m$.

To suppress numerical heating that plagues many published simulations, all of our simulations employ a fifth-order interpolation scheme and exactly conserve total momentum by removing common self-forces on particles \citep{sharp,sharp2}. Since numerical heating accumulates secularly over time \citep{hockney-eastwood}, we report the energy error at the end of the run as an upper bound over the full evolution. At the end of our $5\lambda_m$ PIC simulation ($t\,\Omega_i \simeq 600$, where $\Omega_i=qB_0/m_i$ is the ion cyclotron frequency and $q$ is the ion charge), the total energy changes only by $\sim 10^{-4}\%$ of its initial value, or $0.011\%$ of the initial CRi kinetic energy -- the free-energy reservoir driving the instability. All FPIC simulations show an even smaller energy error, as expected, since the background species evolve as fluids rather than discrete and, therefore, noisy macroparticles.

\begin{figure*}
\includegraphics[width=1.00\linewidth]{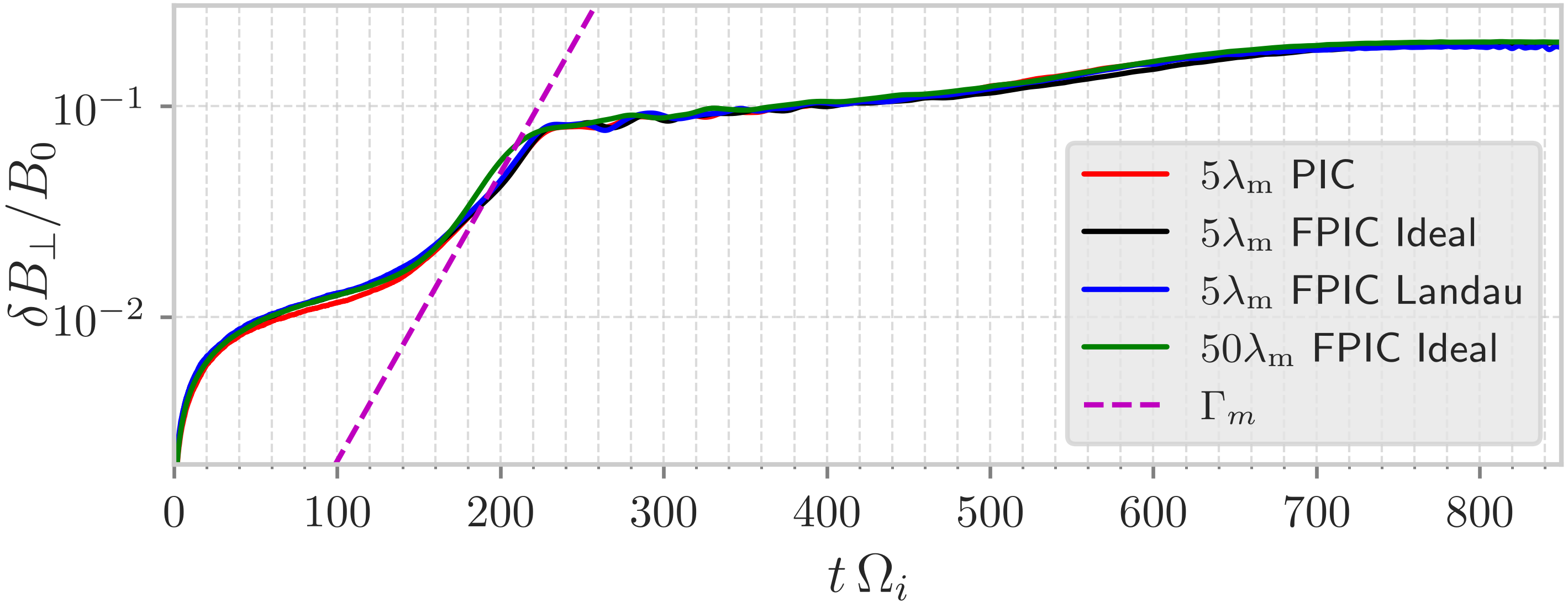}
\caption{\label{fig02}%
Growth of the magnetic-field energy in various simulations. The dashed pink line marks the linear growth rate of the fastest-growing forward-propagating driven {\alf} wave mode, predicted by $\Gamma_m = 0.0314 \Omega_i$. This rate is in excellent agreement with the early-time growth of the total magnetic energy in all simulations. All simulations saturate eventually at $\delta B/B_0 \approx 0.25$. The curves labeled $5\lambda_m$ represent a domain size of $5$ times the fastest-growing wavelength. The green curves correspond to the FPIC simulation with ideal closure, for which the domain extends to $50$ times the fastest-growing wavelength. The close agreement between PIC and FPIC with both closures demonstrates that NLLD is not active at our parameters.}
\end{figure*}

\section{Simulation results}
\label{sec:results}

This section presents the main results of our numerical investigation. We begin by examining the growth of magnetic-field energy across all simulations, then analyze the helicity decomposition of the driven waves, and finally identify the saturation mechanism through various diagnostics. Additional animations showing the full evolution of the various simulations are provided at \citet{Shalaby2026video}.

\subsection{Magnetic field amplification}

Figure~\ref{fig02} shows the growth of magnetic-field energy for all simulations. The dashed pink line indicates the linear growth rate of the fastest-growing mode \citep{Zweibel2017}
\begin{equation}
    \Gamma_m = \frac{\pi}{4} \frac{n_{\rm CR}}{n_i}
    \left( \frac{v_{\rm dr}}{\vA}-1 \right) \Omega_i \simeq 0.0314 \Omega_i,
\end{equation}
which agrees well with the early-time growth in all simulations.
As noted above, the $5\lambda_m$ simulation domain is $L \simeq 5.4\,\lambda_m$, so only discrete wavenumbers $k_n = 2\pi n/L$ are permitted by the periodic boundary conditions \citep{resolution-paper}.
The mode that emerges as the fastest-growing in this domain contains six wavelengths ($n=6$, i.e., $k_n \simeq 1.105\,k_m$), which do not coincide with the true fastest-growing mode at $k_m$. Because the growth rate curve $\Gamma(k)$ falls off away from its peak \citep{Bai2019}, this discretization yields a slightly lower growth rate than in the $50\lambda_m$ simulation, where the fastest-growing mode fits better within the simulation domain ($n=55$, i.e., $k_n \simeq 1.013\,k_m$).

Importantly, the $5\lambda_m$ PIC simulation, the $5\lambda_m$ and $50\lambda_m$ FPIC simulations with ideal closure, and the $5\lambda_m$ FPIC simulation with Landau closure all converge to nearly identical saturation levels, $\delta B/B_0 \approx 0.25$, independent of the simulation domain size and closure type. This agreement demonstrates that the saturation mechanism is local and does not depend on domain size. Furthermore, the FPIC simulation with ideal closure (which treats the background electrons and ions as an ideal fluid, thereby omitting NLLD entirely) matches both the PIC simulation and the FPIC simulation with Landau closure, providing strong evidence that NLLD does not contribute to the saturation at our simulation parameters.

Contrary to the recent claims of \citet{Schroer2025}, the close agreement between the PIC and FPIC-ideal saturation amplitudes -- despite the complete absence of any damping channel in the ideal closure -- already indicates that NLLD does not set the saturation level at our parameter regime; we assemble the full argument for this conclusion in Section~\ref{sec:nlld}. \citet{Lemmerz2026} find consistent closure-agreement at parameters closely matching those of \citet{Schroer2025}, corroborating this picture. Having established the saturation amplitude, we now turn to the mechanism by which this saturation occurs for CRis with a shell momentum distribution, which we investigate below.

\subsection{Unstable Waves}
\label{sec:unstable}

\begin{figure}
    \includegraphics[width=1.0\linewidth]{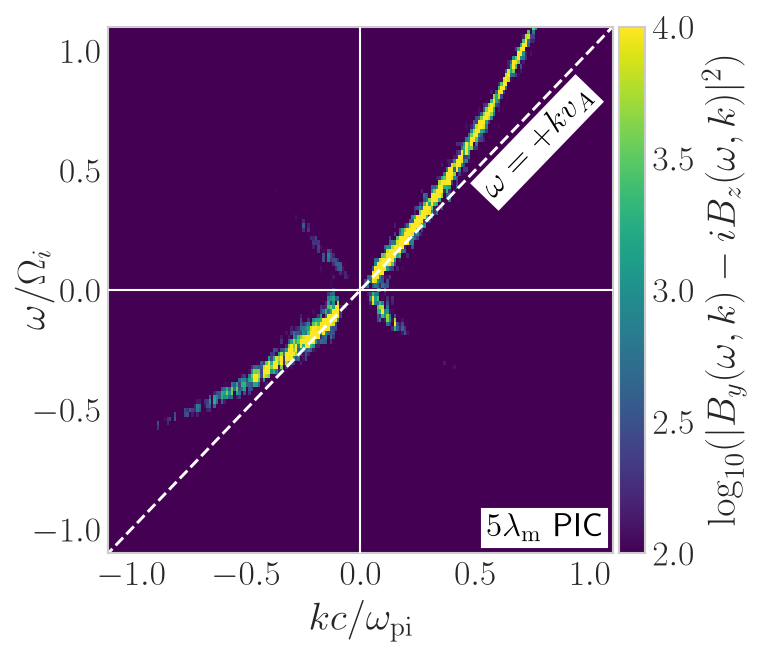}
\caption{\label{fig03}%
Power spectrum of the driven wave modes at the $5\lambda_m$ PIC simulation. The spatial Fourier transform is computed over the full periodic domain, and the temporal inverse Fourier transform is taken over the time interval $t\Omega_i \in [250,600]$, covering both the linear regime and the subsequent nonlinear phase. Most of the spectral power resides in forward-propagating {\alf} waves, with right-handed modes ($k>0$; electron-like gyration) and left-handed modes ($k<0$; ion-like gyration) excited during distinct epochs of the evolution (see Figure~\ref{fig04}). Towards scales of $\left|k c/\omega_\mathrm{pi}\right|\approx 1$, the waves become increasingly dispersive and deviate from the Alfv\'en wave dispersion of $\omega=k\vA$ as expected in kinetic theory. }
\end{figure}

The CRi population has a net drift along $\bs{B}_0$, however, the shell distribution also contains particles with negative parallel velocities in the background frame (see Figure~\ref{fig:CRi}). To determine which wave modes are destabilized by this distribution, we consider the pitch-angle cosine $\mu^{'` }$ distribution in the background frame. Starting from a uniform distribution in the CR rest frame, $f(\mu') = 1/2$, the pitch-angle distribution in the background frame $g(\mu) = f(\mu^{'}) \partial_{\mu}\mu^{'}$ is
\begin{eqnarray}
g(\mu) 
&=&
\frac{\left(2 \mu ^2-1\right) v_r^2+2 \mu  v_r \sqrt{\left(\mu ^2-1\right) v_r^2+1}+1}{2 \sqrt{\left(\mu ^2-1\right) v_r^2+1}}
\qquad 
\end{eqnarray}
where $v_r \equiv v_{\rm dr}/v_0$ and $v_0$ is the speed of the CRi in their rest frame.
For the parameters of our simulations ($v_r \simeq 0.38$), $\partial_\mu g(\mu) > 0$ for all $\mu \in [-1,1]$.

The growth rate of the unstable wave modes can be obtained by solving the dispersion relation, which includes contributions from both the background plasma and the CR species \citep{Kulsrud1969,Zweibel2003,Zweibel+Everett_2010,sharp2,Holcomb2019}. In the limit $\alpha \equiv n_{\rm CR}/n_i \ll 1$, the growth rates simplify considerably; for example, Equations (4) and (5) of \citet{Zweibel2017} provide expressions for the growth rate of the resonant instability in such a limit. While these expressions were originally derived for a power-law CR distribution, the same condition applies to our shell distribution: unstable modes for the case with $\partial_\mu g(\mu) > 0$ require a positive phase velocity $v_\mathrm{ph} = \omega/k > 0$. We therefore expect only forward-propagating modes (those with $v_\mathrm{ph} > 0$) to be driven unstable in our simulations.

Figure~\ref{fig03} shows the power spectrum of the driven wave modes in the $5\lambda_m$ PIC simulation. The spatial Fourier transform is computed over the full periodic domain, and the temporal inverse Fourier transform is applied over the time interval $t\,\Omega_i \in [250,600]$, covering both the linear regime and the subsequent nonlinear evolution. The spectrum reveals that most of the spectral power is concentrated in forward-propagating {\alf} waves ($\omega/k > 0$), with right-handed modes ($k>0$; electron-like gyration) and left-handed modes ($k<0$; ion-like gyration) being excited during the evolution. This confirms our theoretical expectation that only forward-propagating modes are driven unstable by the shell distribution in our simulations.
Because the resonance condition, $k (v_\mathrm{cr} \mu -\vA)=\Omega_i/\gamma_\mathrm{cr}$, maps each value of $\mu$ to a distinct wavenumber $k$ at a fixed velocity $v$ and CR Lorentz factor $\gamma_\mathrm{cr}$, with the sign of $\mu$ determining the helicity (i.e., the sign of $k$) of the driven wave, the observed spectrum reflects contributions from many $\mu$-resolved resonances. The distinct spectral signatures of the two helicities provide the first indication that the saturation mechanism involves individual CRi populations that are resonant with each wave, which we examine in detail below through helicity-resolved wave growth and gyro-phase diagnostics.

\subsection{\protect{\alf} Wave-frame isotropization: Saturation stages in simulations}
\label{sec:saturation}

\begin{figure*}
\includegraphics[width=1.0\linewidth]{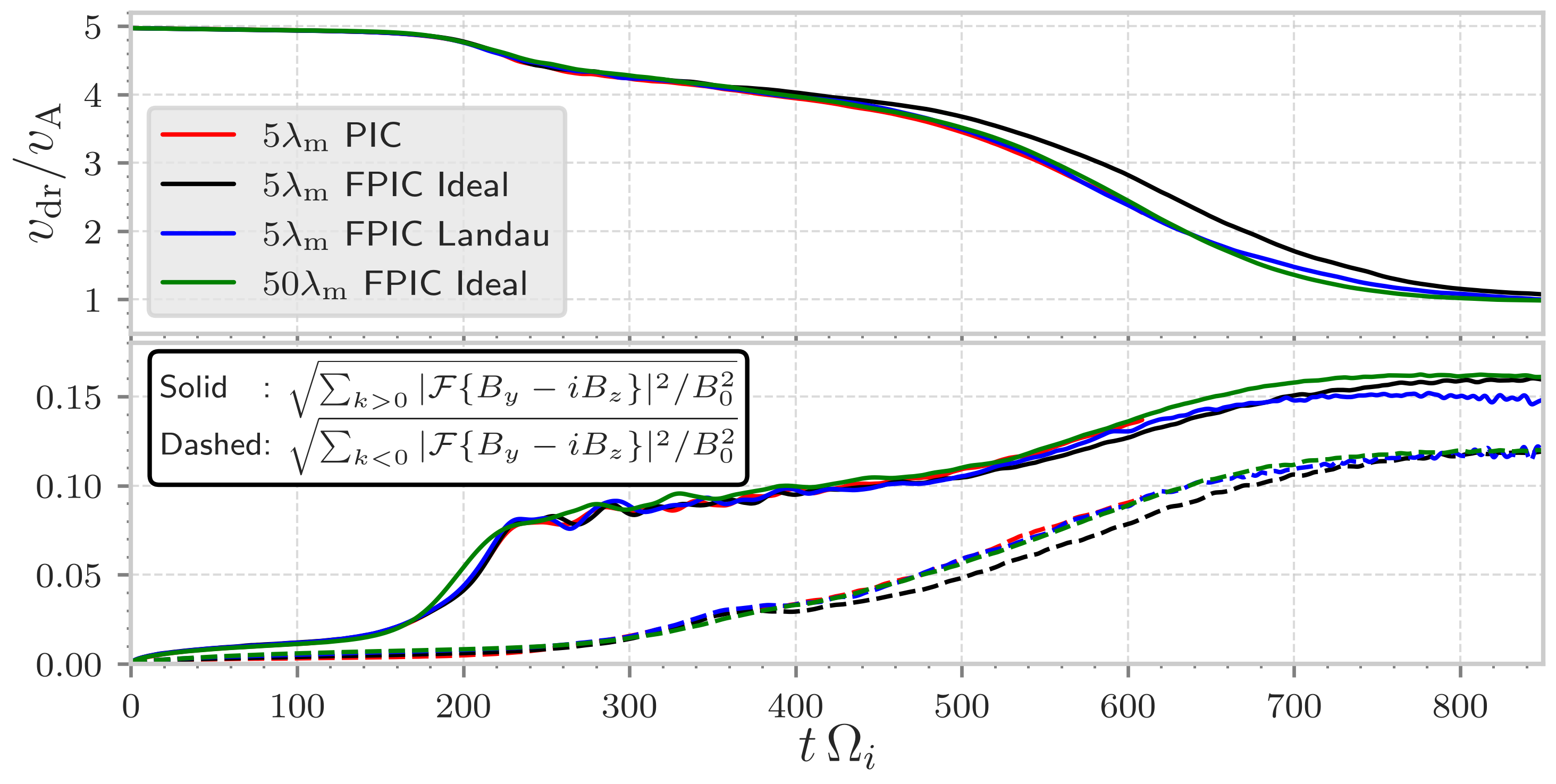}
\caption{\label{fig04}%
\textit{Top panel:} evolution of the average CRi drift velocity, normalized by the initial {\alf} speed. \textit{Bottom panel:} time evolution of the magnetic field amplification in driven {\alf} waves.
Solid lines denote the field amplitude in right-handed polarized waves (electron-like gyration), while Dashed lines denote the field amplitude in left-handed polarized waves (ion-like gyration). Both helicities correspond to forward-propagating modes ($\omega/k > 0$; see also Figure~\ref{fig03}).
    }
\end{figure*}

The top panel of Figure~\ref{fig04} shows the evolution of the average CRi drift speed (normalized by $\vA$) across all simulations, again demonstrating excellent agreement among the different numerical approaches. To quantify the helicity of the driven waves, we define the square of the right-handed (RP) and left-handed (LP) polarization wave amplitudes as
\begin{eqnarray}
    b_\mathrm{RP \,(LP)}^2 &\equiv&
    \sum_{k>0 ~\,(k<0)} \frac{|\mathcal{F} \{ B_y - i B_z\}|^2}{B_0^2}
\end{eqnarray}

The bottom panel of Figure~\ref{fig04} shows the evolution of these helicity-resolved wave amplitudes. The evolution proceeds through two distinct phases. During the first phase ($t\, \Omega_i \in [150,250]$), right-handed modes ($k>0$) grow rapidly, scattering the CRs and reducing their drift velocity from $5\vA$ to approximately $4\vA$. During the second phase ($t\, \Omega_i \in [450,800]$), both helicities grow at a slower rate, driving more efficient CR scattering and further reducing the mean drift velocity to $\vA$.

The growth of both helicities is a direct consequence of the shell distribution: CRs with pitch angles $\mu \gtrapprox 0$ excite right-handed waves, while those with $\mu \lessapprox 0$ excite left-handed waves. As the CR distribution evolves, both populations contribute to wave growth at different stages.

\begin{figure*}
    \includegraphics[width=1.0\linewidth]{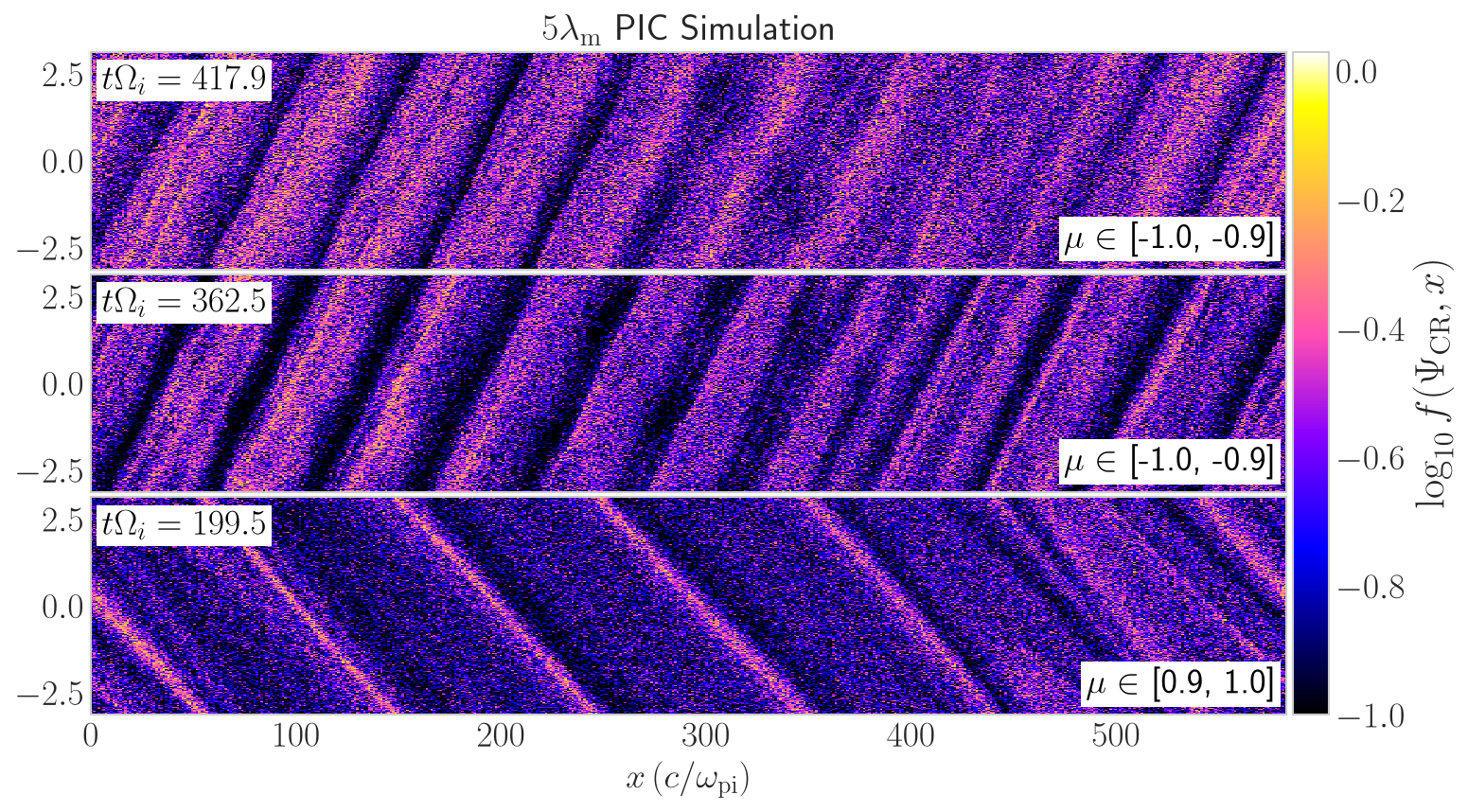}
    \caption{\label{fig04a}%
    Normalized CRi gyrophase distribution (color scale) throughout the simulation domain at three different times during the evolution of the $5\lambda_m$ PIC simulation. The initial CRi gyrophase distribution is uniform over all angles, as shown in \citet{Lemmerz2025}. 
    \textit{Bottom panel:} Gyrophase distribution of CRi with cosine pitch angles $\mu \in [0.9,1.0]$. These particles resonantly drive the right-handed-helicity wave modes. The gyrophase bunching of these CRi around the driven waves demonstrates their resonant interaction with the waves and leads to the saturation of their growth at $t\,\Omega_i \simeq 220$ (see Figures~\ref{fig02} and \ref{fig03}). 
    \textit{Middle panel:} Gyrophase distribution of CRi with $\mu \in [-1,-0.9]$, which resonantly drive the left-handed-helicity wave modes. During the initial phase of wave growth (see the dashed lines in the bottom panel of Figure~\ref{fig04}), these CRi become gyrophase-bunched, leading to the saturation of the corresponding driven waves. 
    \textit{Top panel:} At later stages, the right-handed-helicity waves continue to be driven by CRi with $\mu \in [-1,-0.9]$, with the CRi remaining gyrophase-bunched around the resonantly driven waves. This continued interplay between gyrophase bunching and wave driving persists until the CRi become fully isotropic in the {\alf} wave frame at $t\,\Omega_i \gtrsim 800$, when the instabilities reach their final saturated state.}
\end{figure*}

\begin{figure*}
    \includegraphics[width=1.0\linewidth]{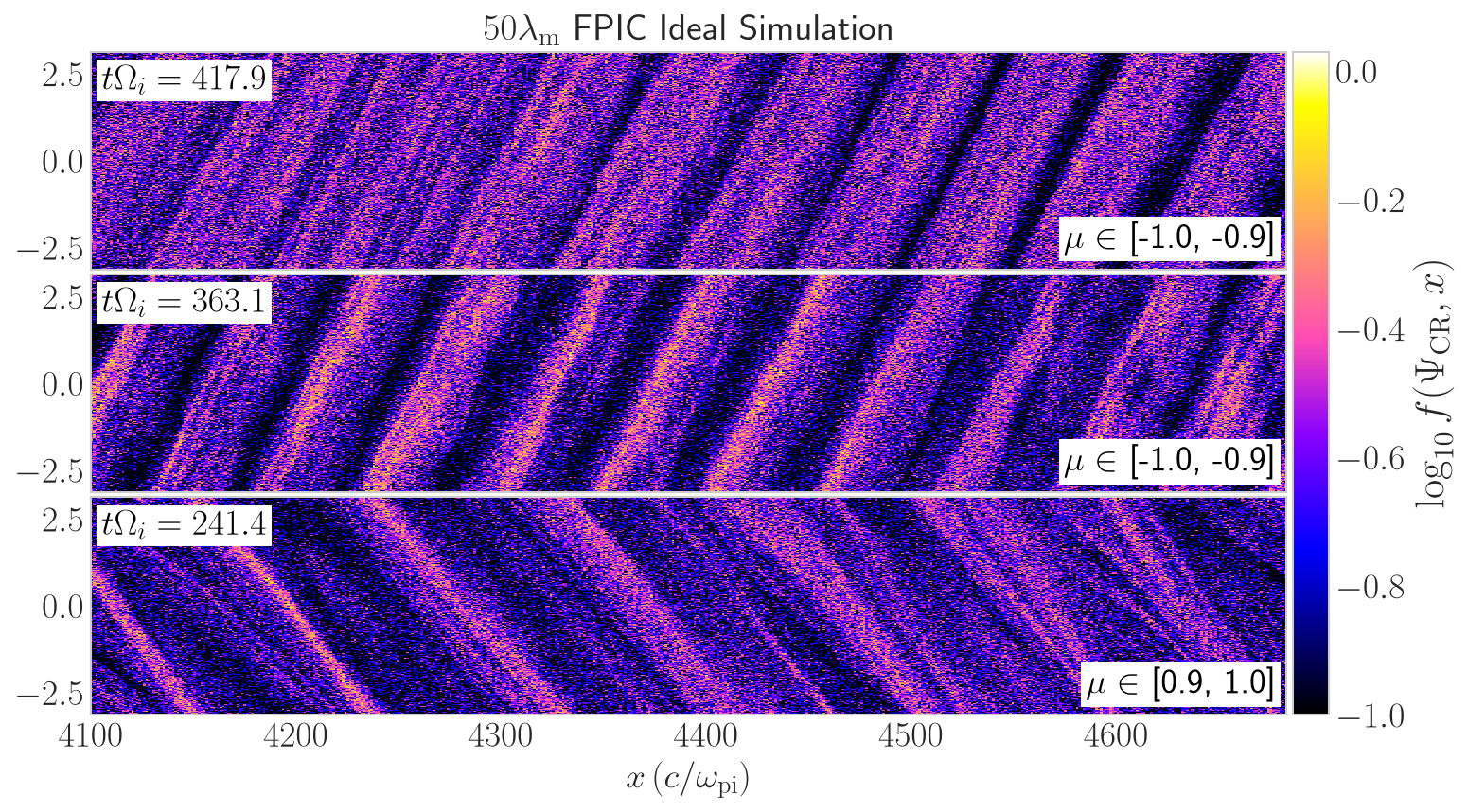}
    \caption{\label{fig04b}%
    Same as Figure~\ref{fig04a} for the $50 \lambda_m$ FPIC simulation. The times shown are slightly different but illustrate the same physical mechanism. In this simulation, the domain extends over $6000 c/\omega_\mathrm{pi}$; we show only a section of the simulation domain. Despite the larger domain and different numerical approach, the gyro-phase bunching mechanism is qualitatively identical to that seen in the PIC simulation (Figure~\ref{fig04a}), demonstrating that the fluid closure captures the essential saturation physics.}
\end{figure*}

To identify the saturation mechanism, we examine the CRi gyrophase distribution in the PIC simulation. We define the gyrophase as the angle, measured from the $y$-axis, of each CR ion's perpendicular velocity vector $\bs{v}_\perp$ in the plane transverse to the background magnetic field $\bs{B}_0 = B_0\bs{\hat{x}}$.
Figure~\ref{fig04a} shows this distribution in the $5\lambda_m$ PIC simulation at three different times. Initially, CRi gyrophases are uniformly distributed over all gyrophase angles. As the instability grows, CRs become gyro-phase bunched around the resonantly driven waves \citep{Lemmerz2025}. The bottom panel of Figure~\ref{fig04a} shows CRi with $\mu \in [0.9,1.0]$, which resonantly drive right-handed modes. These particles exhibit clear gyro-phase bunching at $t\Omega_i \simeq 200$, coincident with the saturation of wave growth (see Figures~\ref{fig02} and~\ref{fig04}). The middle panel shows CRi with $\mu \in [-1,-0.9]$, which resonantly drive left-handed modes. During the initial phase of wave growth, these particles also become bunched, leading to the saturation of the corresponding modes. The top panel demonstrates that, at later times, the interplay between gyro-phase bunching and wave driving continues until the CRs become fully isotropic in the {\alf} wave frame at $t\Omega_i \gtrsim 800$. Thus, the PIC simulation establishes that gyro-phase bunching is the mechanism that saturates the instability at various stages for different $\mu$ values.

Although the temporal coincidence between gyro-phase bunching and the saturation of wave growth might at first suggest only a correlation, the two are causally linked, and the underlying physics is a familiar one. For a given CRi cosine pitch angle $\mu$ and velocity $v_{\rm cr}$\footnote{The CRi velocity in the background rest frame $v_{\rm cr}$ can be related to the CRi $\mu^{'}$ in the CR rest frame as $v_{\rm cr}^2 = v_0^2 \left[ (\mu^{'}+v_r)^2+ 1 - (\mu^{'})^2\right]$, where $v_r = v_{\rm dr}/v_0.$}, the resonance condition
\begin{eqnarray}
    -\Omega_i /\gamma_\mathrm{cr}+ k v_{\rm cri} \mu = k \vA ,
    \label{eq:resonance}
\end{eqnarray}
selects a corresponding wave mode $k$; as its amplitude grows, the wave force on the resonant CRi population becomes significant enough to drive these particles from an initially uniform gyrophase distribution toward a coherent one \citep[the onset of this bunching is documented in][]{Lemmerz2025}.
This gyro-phase bunching is, formally, an instance of resonant particle trapping in a pendulum-type potential. As shown by \citet{Lemmerz2025}, the trapping is driven by the same kind of force as in classical Landau trapping: a force parallel to $\bs{B}_0$, which here arises from the parallel component of the Lorentz force $\bs{v}_\perp \bs\times \vec B_\perp$ sourced by the transverse field of the driven wave.
This parallel force modulates $v_\parallel$, which, through the gyroresonance condition $\omega - kv_\parallel = \Omega_i$, directly controls the rate of change of the CR gyrophase relative to the wave $\varphi \equiv \psi_{\rm cr}-\psi_B$; combining these relations yields a pendulum equation for $\varphi$, with trapped-particle libration replacing free gyration once the wave amplitude exceeds the local trapping threshold. The mechanism is therefore structurally identical to Landau-resonant trapping \citep{ONeil1965}, with the resonant phase variable being the gyrophase $\varphi$ rather than the parallel spatial phase, reflecting the gyroresonance condition in place of the plain Landau resonance.

Saturation follows for the same reason as in parallel trapping: as the resonant CRi population is driven toward $\varphi \approx 0$, the perpendicular CR current that drives the instability is generated. Nearing saturation, the CRs' gyrophases (and the associated currents' phases) oscillate around $\varphi=0$, where the momentum exchange that was driving the wave vanishes and eventually reverses as the average gyrophase (and current) reaches $\varphi \lesssim 0$. The oscillating momentum exchange due to the CRs pendulum motion is observable as periodic ripples in the magnetic amplitude seen in Fig.~\ref{fig02} starting at roughly $t \Omega_i\gtrsim 220$. We derive and characterize this pendulum-trapping structure for the CR gyroresonance in detail in \citet{Lemmerz2025}; here we confirm that it is the saturation channel for the CRSI driven by CRs with a shell-distribution.

To show that this saturation mechanism operates independently of whether NLLD is present in the background plasma, we perform the same diagnostic in the $50\lambda_m$ FPIC simulation with ideal closure. Figure~\ref{fig04b} shows the gyrophase distribution in this simulation at analogous times. Despite the different numerical approach and the larger domain, the gyro-phase bunching behavior is qualitatively identical to that seen in the PIC simulation, consistent with the matching saturation amplitudes already shown in Figure~\ref{fig02}. Minor differences in the detailed structure of the bunched phases arise from the different domain sizes and from the fact that the FPIC simulation treats the background as a fluid. This agreement establishes that the fluid closure faithfully reproduces the microscopic saturation mechanism identified in the fully kinetic treatment.

\subsection{Phase space distribution at nonlinear stages}

\begin{figure*}
    \includegraphics[width=1.0\linewidth]{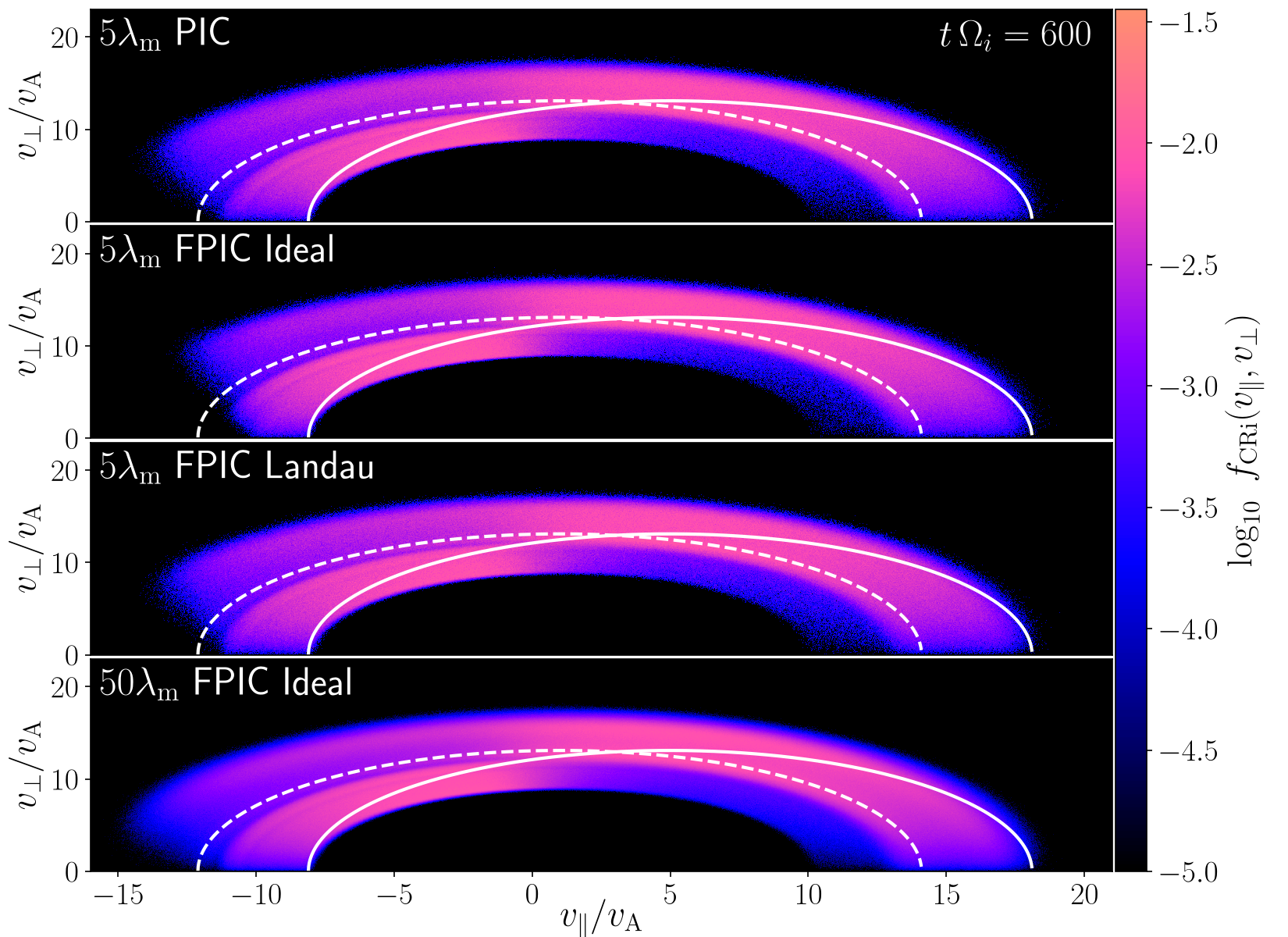}
    \caption{CRi phase space distribution at $t \, \Omega_i = 600$, deep in the nonlinear regime for all simulations. The solid white line indicates the initial distribution of CRi, while the dashed white line indicates the initial distribution shifted to the local {\alf} frame. The shift toward the {\alf} frame confirms that CRs are in the process of being isotropized in the wave frame.}
    \label{fig07}
\end{figure*}

Figure~\ref{fig07} shows the phase space distribution of CRi at $t\Omega_i = 600$, deep in the nonlinear regime for all simulations. The solid white line indicates the initial shell distribution of CRi, while the dashed white line indicates the same distribution shifted to the local {\alf} frame. At this time, the CRi distribution has shifted significantly from its initial shell toward the {\alf} frame, consistent with the reduction in mean drift velocity from $5\vA$ to approximately $\vA$ seen in the top panel of Figure~\ref{fig04}.

Crucially, the phase space distributions from the PIC and FPIC simulations are qualitatively indistinguishable, providing additional support that the FPIC framework with the ideal gas closure captures the essential physics of the nonlinear evolution. This extends previous simulations, where FPIC has been compared against PIC with a simplified, gyrotropic ring CR streaming setup \citep[][\S 3.5]{Lemmerz2023}. There, FPIC with the ideal gas closure correctly captured the evolution of excited {\alf} waves as well. 

The shift toward the {\alf} frame confirms that CRs are being isotropized in the wave frame, which leads to the phase of the gyro-phase bunching mechanism, as discussed above. This isotropization, together with the gyro-phase bunching observed in Figures~\ref{fig04a} and~\ref{fig04b}, establishes a complete picture of the saturation process: CRs become phase-locked to the resonantly driven waves, limiting further wave growth, and are gradually scattered until they isotropize in the {\alf} frame.

\subsection{Irrelevance of NLLD}
\label{sec:nlld}

A central result of this work is the demonstration that NLLD does not contribute to the saturation of the CR streaming instability for the parameters considered here. The evidence for this conclusion rests on a controlled comparison between three simulation approaches that differ only in how they treat the background plasma:

\begin{itemize}
    \item The \textbf{fully kinetic PIC simulation} treats both the background species (electrons and ions) and the CRs as kinetic particles. This simulation includes NLLD self-consistently through the full Vlasov–Maxwell system, without any approximation. As such, comparisons involving this simulation carry the primary evidential weight in what follows, since it makes no assumptions about the strength or form of NLLD.
    \item The \textbf{FPIC simulation with ideal closure} treats the background electrons and ions as a charged, ideal fluid with no dissipation. This closure omits NLLD entirely, as well as any other wave–particle damping processes in the background plasma.
    \item The \textbf{FPIC simulation with Landau closure} retains the same fluid treatment for the background but includes a closure that approximates the effects of linear and thus nonlinear Landau damping on the background species \citep{Lemmerz2023}. This closure has been validated against theoretical NLLD predictions in the interacting-{\alf}-wave test of \citet[][\S 3.4]{Lemmerz2023} to good accuracy.
\end{itemize}
That is, the agreement between the FPIC-Landau and PIC simulations provides supporting, rather than decisive, evidence regarding NLLD: a match could, in principle, arise if NLLD were weakly active in the PIC simulation but too weak for the closure to resolve. The comparison that is not subject to this ambiguity is between the PIC simulation and the FPIC simulation with ideal closure, since the latter contains no damping channel whatsoever.

If NLLD were an important saturation mechanism, we would expect the ideal closure simulation to diverge from the PIC simulation, with the ideal closure overestimating the wave amplitude because it lacks any damping channel. Instead, we find that \textit{all three simulations produce closely matching saturation amplitudes} ($\delta B/B_0 \approx 0.25$) and, crucially, exhibit the \textit{same gyro-phase bunching mechanism} (Figures~\ref{fig04a} and~\ref{fig04b}). The agreement is not limited to the saturation level; it holds throughout the entire evolution, from the linear growth phase through the nonlinear stages (Figure~\ref{fig02}). Because the PIC vs.\ FPIC with ideal--closure comparison alone already rules out a role for any damping channel, including NLLD, this indicates that NLLD is dynamically irrelevant not only at saturation but also during the preceding nonlinear development.

The physical reason for this irrelevance is that the saturation is governed by a different process: gyro-phase bunching of the CRs around the resonantly driven {\alf} waves. Once the CRi become phase-locked to the waves, the wave--particle energy exchange saturates regardless of whether the background plasma can damp the waves. The background damping channels, including NLLD, are simply not fast enough or strong enough to compete with the CR-driven growth and the subsequent phase bunching. The ideal closure, which omits NLLD, captures the same dynamics because the waves are not limited by background dissipation but by the CR response itself.

We emphasize that our conclusion applies to the parameters investigated here: $\vA = 0.01c$, $n_{\rm CR}/n_i = 0.01$, $v_{\rm dr}/\vA = 5$, and a shell momentum distribution with $v_0 = 13.1\vA$. Extrapolation to different parameter regimes -- particularly to the much lower CR densities, {\alf} speeds, and higher plasma $\beta$ typical of the interstellar medium — requires further investigation. However, the agreement between PIC and FPIC with ideal closure provides strong evidence that, at least in the regime where the instability is vigorous, NLLD is not the bottleneck for saturation. 
This conclusion is consistent with, though not independently confirmed by, \citet{Lemmerz2026}, who performed FPIC simulations with both ideal and Landau closures at parameters closely matching those of \citet{Schroer2025}: the two closures continue to agree there. However, since that comparison lacks a fully kinetic PIC benchmark, it corroborates rather than establishes the irrelevance of NLLD in that regime. Together, these results challenge the common assumption in CR-MHD models that NLLD is a dominant saturation mechanism for the streaming instability \citep{Jiang2018,timon2019}.

\section{Discussion}
\label{sec:discussion}

Our results have important implications for the modeling of CR transport in astrophysical systems. The identification of gyro-phase bunching as the saturation mechanism, rather than NLLD, suggests that two-moment CR-MHD models that rely on NLLD as the dominant saturation mechanism may require revision, at least in the parameter regime relevant to our simulations.

\subsection{Comparison with {\protect \citet{Bai2019}}}

The MHD-PIC method is employed by \citet{Bai2019} to study the growth and saturation of the CR streaming instability in a regime closer to realistic interstellar medium (ISM) conditions, with parameters such as $n_{\rm CR}/n_{\rm ISM} \sim 10^{-4}$ and $\vA/c \sim 3 \times 10^{-3}$.
They captured the full evolution of the instability from linear growth to saturation in their simulations.

The most significant differences between our work and \citet{Bai2019} lie in the methodology and the physical mechanisms on which we focus. We highlight three key aspects:

\begin{itemize}
    \item \textbf{Different perspective on the saturation mechanism:} \citet{Bai2019} attribute saturation to quasi-linear diffusion, noting that full isotropization of CRs in the wave frame requires efficient crossing of the $90^\circ$ pitch angle, which they attribute to nonlinear wave-particle interactions rather than mirror reflection. A key mechanism enabling this diffusion in their simulations is phase randomization across periodic boundaries.

    \item \textbf{Our findings:} In contrast, our fully kinetic PIC simulations reveal a more microscopic picture of the saturation. We find that the instability saturates through gyro-phase bunching around the driven {\alf} waves. This bunching occurs simultaneously with the wave modes reaching their maximum amplitude, thereby limiting the efficiency of wave-particle energy exchange.

\end{itemize}

\subsection{Comparison with \protect \citet{timon2019}}

Our results suggest an extension of the CR-MHD two-moment models developed by \citet{timon2019}. In those models, the evolution of {\alf} waves is governed by a balance between CR-driven excitation and various damping processes. Among these, NLLD is included as a key dissipation channel, whereby wave energy is transferred to the background thermal plasma via resonant interactions with thermal particles \citep[see, e.g.,][]{Miller1991,Skilling1975c}. Within this framework, the NLLD damping rate scales linearly with the wave intensity, $e_{a,\pm}$, implying an energy damping rate \( \propto e_{a,\pm}^2\) that becomes increasingly efficient at high wave amplitudes. This would, in turn, predict a reduced saturation level of \(\delta B/B_0\) compared to models that omit such damping. Thus, in the two-moment picture, NLLD acts as a key regulator of the CR--wave interaction, limiting both wave growth and the efficiency of CR scattering.

Our simulations, however, show that the saturation amplitude \(\delta B/B_0 \approx 0.25\) is independent of whether NLLD is included (PIC), omitted (FPIC ideal), or approximated (FPIC Landau). This indicates that, at least at our parameters, NLLD does not act as a regulator. The saturation is instead governed by the local gyro-phase bunching of CRs around the driven waves, which limits wave growth regardless of background damping channels. This does not necessarily invalidate the two-moment models in all regimes, but it suggests that the parameter regime relevant to our simulations  --  where the instability is vigorous and the CR density is relatively high  --  is not dominated by NLLD.
We leave to future work the question of whether this conclusion holds at the lower CR densities and {\alf} speeds typical of the interstellar medium.

\subsection{Comparison with \protect\citet{Schroer2025}}

Recent work by \citet{Schroer2025} using hybrid-PIC simulations has claimed that NLLD plays an important role in the saturation of the CR streaming instability, reducing the CR drift speed through scattering while keeping it super‑{\alf}ic owing to the damping of the generated waves~\citep{Miller1991}.
Our results directly challenge this claim, at least for the parameters investigated here. While \citet{Schroer2025} argue that NLLD suppresses wave growth and reduces the saturation amplitude, our controlled comparison between PIC, FPIC ideal, and FPIC Landau simulations shows no such suppression. The discrepancy may arise from differences in the numerical methods (and the associated sources of numerical noise), the specific parameter regimes considered (e.g., the CR density and {\alf} speed), or the different treatments of the background plasma (hybrid‑PIC vs.\ fully kinetic or fluid‑PIC). A detailed comparison of the two approaches is beyond the scope of this work, but our results suggest that the role of NLLD may be more subtle than previously assumed and may depend sensitively on the physical parameters of the system.

\subsection{Implications for subgrid CR transport models}

The domain-size independence of the saturation amplitude indicates that the saturation mechanism is local and does not require global-scale phase randomization. The gyro-phase bunching occurs at the resonant wave-particle scale, not at the domain scale. This has direct implications for subgrid models of CR transport.

Many subgrid models assume that the saturation amplitude of the streaming instability is determined by global properties of the system, such as the CR pressure gradient or the {\alf} speed, and often include NLLD as a key damping process. Our results suggest that, at least in the regime where the instability is driven by a shell-like CR distribution, the saturation amplitude is set locally by the resonant wave-particle interaction. This implies that subgrid models may need to account for the local microphysics of gyro-phase bunching rather than relying solely on global damping processes.

The saturation amplitude $\delta B/B_0 \approx 0.25$ found in our simulations provides a concrete benchmark for subgrid models. This amplitude is unlikely to be universal across different CR density contrasts $\alpha$ and plasma parameters, and thus remains an open question that warrants further investigation. Nevertheless, the robustness of this value across PIC and FPIC simulations with both closures and different domain sizes suggests that it may be a characteristic feature of the instability in the vigorous growth regime.

\subsection{Synthesis}
Taken together, our results establish that the gyro-phase bunching already observed to saturate the much simpler gyrotropic ring distribution \citep{Lemmerz2025}, is also the saturation mechanism for a more realistic shell CR distribution for the  parameter regime investigated here.
This finding suggests an extension of the saturation mechanisms used in popular two-moment CR-MHD models \citep{Jiang2018,timon2019} and challenges recent claims \citep{Schroer2025}. While our results do not rule out NLLD in other parameter regimes, they demonstrate that the saturation mechanism is not universal and depends sensitively on the CR distribution, density, and other plasma parameters. Future work should explore the transition between the gyro-phase bunching-dominated regime identified here and regimes where NLLD might become relevant.

\section{Conclusions}
\label{sec:conclusions}

This paper establishes two principal results.

First, our fully kinetic PIC simulation shows excellent agreement with FPIC simulations using both ideal and Landau closures. The FPIC simulation with ideal closure omits NLLD entirely, yet it matches both the PIC simulation and the FPIC simulation with Landau closure in all respects: the linear growth rate, the saturation amplitude, the helicity-resolved wave evolution, and the gyro-phase bunching mechanism. 
We therefore conclude that NLLD does not affect the saturation of the CR streaming instability for the parameters explored here ($\vA = 0.01\,c$, $n_{\rm CR}/n_i = 0.01$, shell distribution with $v_{\rm dr}/\vA = 5$ and $v_0 = 13.1\vA$). The same FPIC ideal-vs.-Landau agreement persists in \citet{Lemmerz2026} at parameters closely matching those of \citet{Schroer2025}, corroborating, though not independently confirming, this conclusion in that regime, since a fully kinetic PIC benchmark is not available there.

Second, and more significantly, all simulations converge to the same saturation amplitude ($\delta B/B_0 \approx 0.25$) and, crucially, to the same saturation mechanism: gyro-phase bunching around the driven {\alf} waves. This mechanism is local, as demonstrated by the agreement between $5\lambda_m$ and $50\lambda_m$ domains.
The gyro-phase bunching occurs simultaneously with the wave modes reaching their maximum amplitude, limiting the efficiency of wave-particle energy exchange and saturating the instability. The phase-space distributions confirm that CRs are isotropized in the {\alf} wave frame, completing the saturation picture.

The identification of gyro-phase bunching as the saturation mechanism -- rather than NLLD -- carries important implications for CR transport modeling. In particular, two-moment CR-MHD models that assume NLLD as the dominant saturation process may require revision, at least in the parameter regime relevant to our simulations. Future work should investigate whether this conclusion holds at the lower CR densities and {\alf} speeds typical of the interstellar medium, and whether the transition to an NLLD-dominated regime occurs at more realistic parameters.

\section*{Acknowledgments}
This research was supported in part by Perimeter Institute for Theoretical
Physics. Research at Perimeter Institute is supported by the Government of
Canada through the Department of Innovation, Science and Economic
Development and by the Province of Ontario through the Ministry of Colleges
and Universities. M.S. receives additional support through the Horizon AstroPhysics Initiative (HAPI), a joint venture of the University of Waterloo and Perimeter Institute for Theoretical Physics. CP and TT acknowledge support from the European Research Council under the ERC-AdG grant PICOGAL-101019746. This work was supported in part by the National Science Foundation under Grant No. NSF PHY2409224 (RL).
This work was supported by the North-German Supercomputing Alliance (HLRN) under project bbp00078.
This research was enabled in part by the support provided by Calcul Qu\'ebec (\href{www.calculquebec.ca}{www.calculquebec.ca}) and the Digital Research Alliance of Canada (\href{alliancecan.ca}{alliancecan.ca}).

\bibliography{refs}
\bibliographystyle{aasjournal}

\end{document}